\documentclass[%
 reprint,
 amsmath,amssymb,
 aps,
]{revtex4-1}

\usepackage{graphicx}
\usepackage{dcolumn}
\usepackage{bm}
\usepackage{lipsum}
\usepackage{array}
\newcolumntype{L}[1]{>{\raggedright\let\newline\\\arraybackslash\hspace{0pt}}m{#1}}
\newcolumntype{C}[1]{>{\centering\let\newline\\\arraybackslash\hspace{0pt}}m{#1}}
\newcolumntype{R}[1]{>{\raggedleft\let\newline\\\arraybackslash\hspace{0pt}}m{#1}}

\begin{document}


\title{A Survey of Exoplanetary Detection Techniques}

\author{Jason W. Wei}
 \altaffiliation{Dartmouth College}

\date{\today}

\begin{abstract}
Exoplanets, or planets outside our own solar system, have long been of interest to astronomers; however, only in the past two decades have scientists had the technology to characterize and study planets so far away from us. With advanced telescopes and spectrometers, astronomers are discovering more exoplanets every year. The two most prolific methods for detecting exoplanets, radial velocity, which measures the wobble of an exoplanet's parent star, and transit, which detects the passage of a planet in front of its star, are responsible for the discovery of the vast majority of exoplanets we know about so far. Other methods such as direct imaging, timing, and gravitational microlensing are less applicable but can sometimes yield accurate data to confirm planets found by radial velocity or transit photometry. Astronomers will continue to find large numbers of new exoplanets in the near future, but it will become increasingly harder to do so after all the planets close to our solar system are discovered.
\end{abstract}

\pacs{Valid PACS appear here}
\maketitle


\section{\label{sec:level1}Introduction}

Astronomers have been observing exoplanets throughout the past two decades, searching for life elsewhere in the universe. Since exoplanets are very far away, they cannot be directly seen with most telescopes; thus, scientists have devised indirect methods for discovering and characterizing these planets. Section IIA details methods for detecting exoplanets. Section IIB analyzes the advantages and drawbacks of each of these methods. Section IIC is a case study of Kepler-186f. Section III concludes the paper and touches upon future research of exoplanets.

\section{\label{sec:level1}Analysis of Exoplanetary Detection Methods}

In the body of this survey, a summary of exoplanetary detection methods will first be presented. Then, the advantages and drawbacks of each method will be examined in order to to shed light on how astronomers study exoplanets. Finally, we will look at the discovery of Kepler-186f, the first discovered Earth-sized planet in the habitable zone.

\subsection{\label{sec:level2}Exoplanetary Detection Techniques}

Astronomers have been searching for exoplanets long before the first exoplanet B1257+12 was discovered in 1992 [Wolszaczan, 1994]. Since then, they have derived various methods of detecting such planets, with radial velocity and transit methods as the two most prolific. Other methods such as direct imaging, timing, and gravitational lensing are not as common, but have unique advantages and play important roles in the search for exoplanets. Astronomers use one or a combination of these methods to study the mass, size, orbit, temperature, and atmosphere of celestial bodies in the search for a habitable exoplanet.

\subsubsection{The Radial Velocity Method}

The radial velocity (Doppler spectroscopy) method is one of the earliest methods of exoplanetary discovery, with scientists using it to discover a significant number of planets since 1988 [Lovis and Fischer, 2010]. It uses the stellar wobble, or the small orbit of a star around its star-planet center of mass, to search for exoplanets orbiting around the star. This wobble, though very small, can measured as a variation in radial velocity by modern spectrometers such as the High Accuracy Radial Velocity Planet Searcher (HARPS) spectrometer at the European Southern Observatory (ESO) in the La Silla Observatory in Chile, which has observed more than 451 wobbles during its Guaranteed Time Observations (GTO) planet search program [Sousa et al., 2008]. By looking at this wobble, scientists can observe the velocity of the star $v_{\star}$ and period of the orbit \textit{T}. With an estimate of the mass of the star $M_{\star}$ by spectral type [Kornreich, 2015] and the inclination of the orbit $i_{orbit}$ by stellar photospheric absorption lines [Cochran and Hatzes, 1996], the velocity of the star can be expressed as such:

\begin{equation}
v_{\star} = \frac{m_{p} sin(i_{orbit})}{M_{\star}+m_{p}}\sqrt{\frac{G(M_{\star} + m_{p})}{a}}
\end{equation} 
where $m_{p}$ is the mass of the planet and \textit{a} is the radius of an assumed circular orbit. The period is also given by Kepler's third law of planetary motion:

\begin{equation}
T = \frac{4\pi^{2}a^{3}}{G M_{sys}}
\end{equation} 
where the mass of the system $M_{sys}$ = $M_{\star}$ + $m_{p}$. The orbit radius of the planet is unknown, but can be solved for by combining \textit{Equation 1} and \textit{Equation 2}: 

\begin{equation}
v_{\star} = \left(\frac{2\pi G}{T}\right)^{1/3} \frac{m_{p} sin(i_{orbit})}{(M_{\star}+m_{p})^{2/3}}
\end{equation} 
where plugging $v_{\star}$ back into \textit{Equation 1} will allow us to also solve for the orbital radius.

As such, scientists can indirectly find the mass of an orbiting planet. Scientists have been highly successful detecting planets with the radial velocity method, especially when the mass of the planet was large with respect to the mass of the star, creating a easily detectable stellar wobble. However, there are limitations to this method. Because stellar wobbles are typically small, radial velocities are hard to observe at great distances and are consequently only used to examine relatively nearby stars [Baranne et al, 1996]. Furthermore, false signals are often present when observing multi-planet and multi-stellar systems where the center of mass of the system is unclear and long, continuous observations are required to differentiate stellar bodies. To compensate for high noise to signal ratios, scientists often employ complex periodogram calculations by Fourier analysis to detect the period of planets [Cochran and Hatzes, 1996]. Despite such challenges, the radial velocity method remained the most prolific exoplanetary detection technique until 2014 when it was surpassed by transit photometry [Open Exoplanet Catalogue, 2015]. 

\subsubsection{Transit Photometry}
Since its first use in the mid-2000s, the transit photometry method has now surpassed radial velocity searches in number of new planets discovered [Charbonneau, 2006]. In transit photometry, scientists observe the marginal dimming of a star when its planet passes between the observer and the star. With the change in stellar flux $F_{\star}$ and an estimate of the radius of the star $R_{\star}$ from other measurements, we can determine the radius of a planet $R_{p}$ [Seager and Mallen-Ornelas, 2002]:

\begin{equation}
\frac{F_{\star} - F_{\star, transit}}{F_{\star}} = \frac{\Delta F_{\star}}{F_{\star}} = \left(\frac{R_{p}}{R_{\star}}\right)^2
\end{equation} 
Over multiple observations, the period T can also be observed, allowing us to calculate the radius of planetary orbit \textit{a} using Kepler's third law:

\begin{equation}
a^3 = \frac{M_{sys} G T^2}{4\pi^2}
\end{equation} 
where the mass of the exoplanet is negligible in relation to the mass of the host star, so $M_{sys}$ = $M_{\star}$. With more extensive analysis of the flux curve through varying spectra of light, scientists are also able to observe the presence of atmospheres that can change the observed radius of a transiting planet due to their abilities to absorb and reflect certain wavelengths of light [Southworth et al, 2017].

Though transit photometry has been highly successful in discovering large numbers of exoplanets through missions such as the \textit{Kepler} mission, which found 340 planetary systems with 851 planets validated to better than the 99\% confidence level [Rowe et al., 2014], it still has limitations. While transit surveys can scan large areas of the sky containing hundreds of thousands of stars at once [Hidas et al., 2005], they can only observe planets with orbits that perfectly intersect the astronomer and the star, thus missing many possible stellar systems that could contain planets of interest. False positives are also common when small variations in stellar brightness occur due to natural phenomenon such as pulsations of red giant branch stars [Kiss and Bedding, 2003], and studies have found that up to 35\% of transit candidates turn out to be false-positives [Diaz et al, 2012]. On balance, the transit photometry method has been used effectively in missions with the 2005 Spitzer Space Telescope [Charbonneau et al, 2005], the French Space Agency's CoRoT [Deleuil et al, 2009], and the \textit{Kepler} mission to discover more than 3,483 confirmed planets in 581 planetary systems [NASA Exoplanet Archive, 2017]. 

\subsubsection{Less Prolific Methods of Planetary Detection}

In addition to the radial velocity and transit photometry methods, there are other mechanisms for detecting exoplanets such as direct imaging, timing, and gravitational microlensing. In direct imaging, scientists examine thermal radiation of exoplanets directly through infrared imaging. Typically, we can only observe especially large, hot planets both close to the solar system and far from their stars, and imaging of Earth-like planets requires high levels of optothermal stability [Brooks et al, 2015]. 
\begin{figure*}
 \center
  \includegraphics[width=\textwidth]{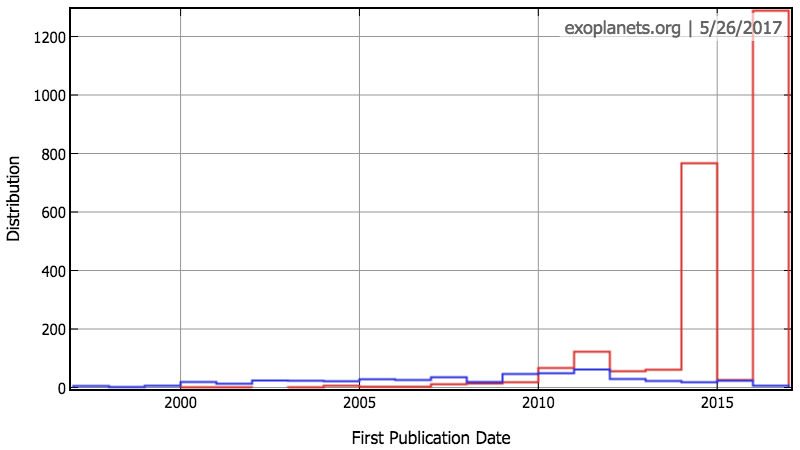}
  \caption{Radial velocity and transit exoplanetary discoveries by year through 2017. Radial velocity, blue; transit photometry, red. In 2014, the number of planets observed by transit photometry surpassed the number of planets observed by radial velocity. Figure created using exoplanets.org.}
\end{figure*}
Timing methods take advantage of niche properties of certain exoplanets and can involve pulsars (neutron stars) that regularly emit radio waves as they rotate [Townsend, 2003], changes in eclipsing binary minima that detect planets far from their host star [Doyle 2011], and transit timing variations caused by interplanetary gravitational pull [Miralda-Escude, 2002]. Because so few exoplanets have these characteristics, this method is also limited in applicability. Finally, gravitational microlensing looks at the marginal effect of a planet on the gravitational lensing of a star behind it. When a foreground star is between the observer and a source star, the foreground star magnifies the light of the source star, with which the planet can make an observable contribution to the lensing effect [Beaulieu et al, 2005]. This method is actually more sensitive for detecting exoplanets with large orbits and is most effective for planets between Earth and the center of the galaxy where there are many background stars, but the alignment needed for gravitational microlensing is rare and cannot be repeated. Though these detection methods are more specialized and are not applicable to as many stars, they have the potential to provide highly accurate data in comparison to the radial velocity and transit photometry methods.

\subsection{\label{sec:level2}Method Comparison and Analysis}

Each method of planetary detection uses different astronomy concepts to deduce properties of exoplanets. While the radial velocity and transit photometry methods have been the most prolific to date (FIG. 1), other methods have also been important in the search for a habitable planet, and each method has its own advantages and disadvantages (TABLE I).
\begin{table*}
\caption{Measurements, advantages, drawbacks, best cases, and instrumentation for the radial velocity, transit photometry, direct imaging, timing, and gravitational microlensing methods of detecting exoplanets. $m_p$, mass of planet; $r_p$, radius of planet; \textit{T}, period of orbit; \textit{a}, length of semi-major axis of orbit.}
\centering
\begin{tabular}{C{0.16\linewidth}C{0.16\linewidth}L{0.16\linewidth}L{0.16\linewidth}L{0.16\linewidth}L{0.16\linewidth}}
\hline
Method & Measurements & Advantages & Drawbacks & Best Case & Instrumentation\\
\hline
\hline
Radial Velocity & $m_p$, \textit{T}, \textit{a} & Can be used to observe many stars & Planet must be close to observer & Small cool star, large planet & Telescope spectrometer\\
Transit Photometry & $r_p$, \textit{T}, \textit{a} & Observe thousands of stars at once, possible atmospheric data & False positives, only observe intersecting planets & Large planet, low noise & Space Observatories (CoRoT and \textit{Kepler})\\
Direct Imaging & $m_p$, \textit{T}, \textit{a} & Detect planets with face-on orbits & Optothermal stability needed, glare & Planet close to observer, large size, large \textit{a}, hot & Infrared telescope\\
Timing & $m_p$, \textit{T} & Detect small planets far from observer, multi- planetary systems, accurate & Few stars & Pulsars, multi- planetary systems, binary star systems & Telescopes, \textit{Kepler}\\
Gravitational Microlensing & $m_p$ & Detect planets with wide orbits & Unlikely alignment, single trial & large \textit{a}, background toward center of galaxy & Robotic telescopes (OGLE)\\
\hline
\end{tabular}
\end{table*}

The radial velocity method is highly prolific and applicable to many planets, though they must be close to the observer. The transit photometry method has observed a large number of stars but has a high rate of false positives. Direct imaging, though only applicable to a small number of stars, finds explicit, unambiguous evidence of exoplanets. Timing methods are similarly limited in applicability, but can provide accurate data of planets that are extremely far away. Gravitational microlensing can detecting planets with large orbital radii, but measurement alignments are unlikely and cannot be repeated. In terms of observed properties, the transit photometry method gives an observed planet's radius, while the other methods provide an indication of the mass of an observed planet.

Though all methods can be used independently, the most accurate searches use more than one technique to gather multi-faceted data on planets of interest. For instance, the radial velocity and transit methods used on the same star can determine mass and radius, yielding density, which hints at the composition of a given planet and its habitability; transit timing can be used to confirm exoplanets in multi-planetary systems discovered by the transit photometry method. Often, discovery of a planet with one method can facilitate observation of its other properties with another.

\subsection{\label{sec:level2}Case Study: Kepler-186f}

A brief case study of Kepler-186f, the first earth-sized exoplanet in the habitable zone, demonstrates the use of planetary detection techniques to characterize exoplanets. Discovered in 2014, Kepler 186f (1.1 $R_{earth}$ and 130-day orbital period) transits Kepler-186, a 0.47 solar-radius star [Quintana et al, 2014]. Originally, transit photometry measurements by the \textit{Kepler} mission only discovered the four planets Kepler-186b to Kepler-186e around Kepler 186, all with short orbital periods between 4 and 22 days (Figure 2a) [Quintana et al, 2014]. However, after another year, scientists collected data on Kepler-186f (Figure 2b) and found that the data fit a five-planet model with limb-darkened transits with high statistical significance by the Markov-chain Monte Carlo (MCMC) algorithm, and calculated properties of Kepler-186f (Table II) [Quintana et al, 2014]. Furthermore, Quintana used planetary thermal evolution models to constrain the composition of Kepler-186f, finding that it has an estimated mass of 1.44 $M_{earth}$ and could contain iron, silicate rock, water, and ice [Quintana et al, 2014]. As such, Kepler is just inside the habitable zone with $32^{+6}_{-4}$\% of the insolation received by Earth from the Sun, and is likely to have properties required to maintain reservoirs of liquid water [Quintana et al, 2014].
\begin{figure}
\center
  (a)\includegraphics[width=0.95\linewidth]{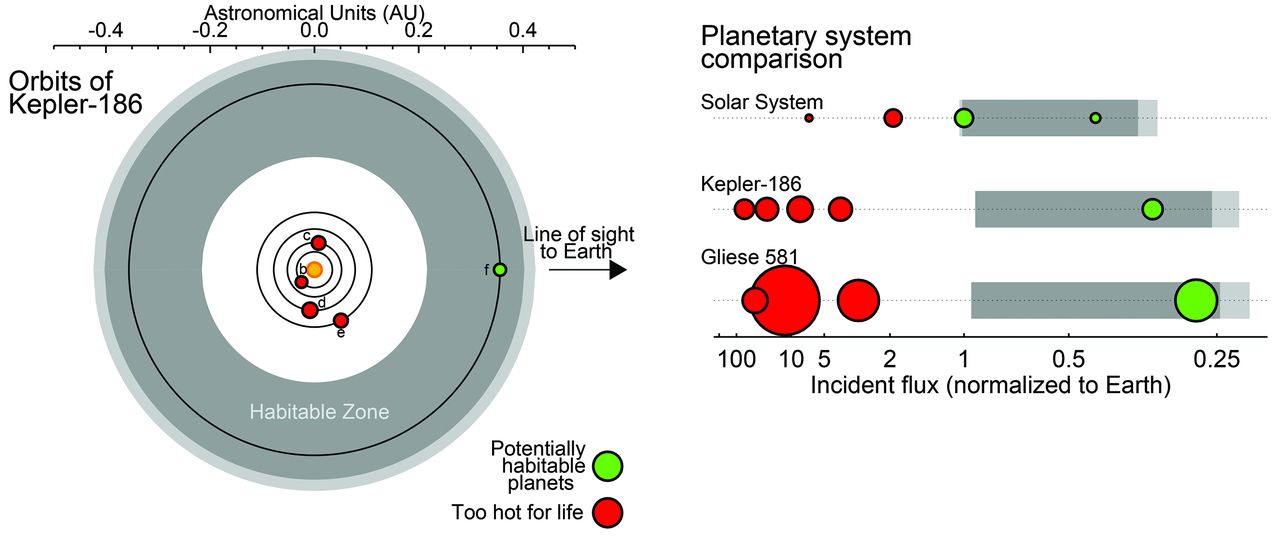}\\ \vspace*{0.3cm}
  (b)\includegraphics[width=0.95\linewidth]{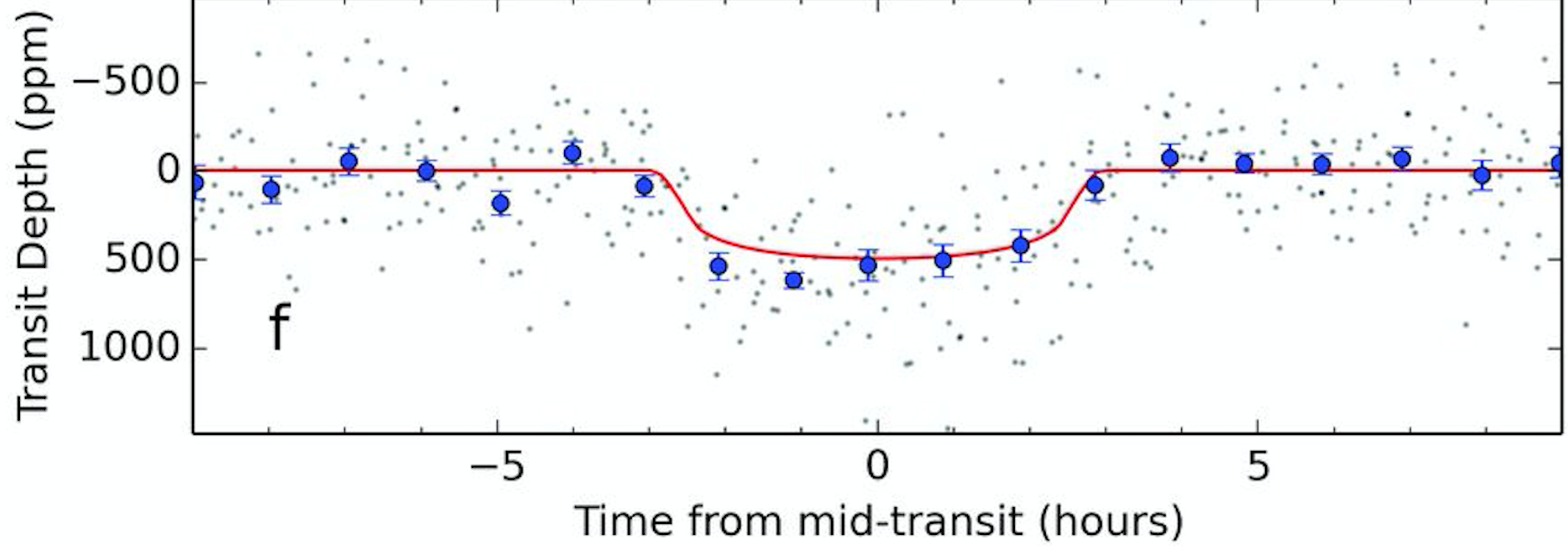}\\
  \caption{Original publication data of Kepler-186f, the first earth-sized exoplanet in the habitable zone. (a) is a schematic diagram of the Kepler-186 system. Dark grey regions show conservative habitable zone estimates, while light grey regions show optimistic habitable zone estimates. (b) shows the transiting planet signal of Kepler-186 as observed by \textit{Kepler}. Black points show observed data, and blue points show observed data binned in time with one point per phase-folded hour. The red line shows the most probable transit model, with Kepler-186f causing a transit dip of approximately 500 ppm. Source: Quintana et al, 2014.}
\end{figure}
\begin{table}
\caption{Properties of Kepler-186f, the first discovered earth-sized exoplanet in the habitable zone.}
\centering
\begin{tabular}{L{0.55\linewidth}C{0.4\linewidth}}
\hline
\hline
Radius & 1.1 $R_{earth}$ \\
Orbital Period & 130 days\\
Mass & 1.44 $M_{earth}$ \\
Insolation & $32^{+6}_{-4}$\% $F_{Sun-Earth}$\\
Habitable & Yes\\
\hline
\end{tabular}
\end{table}

\section{\label{sec:level1}Conclusions and Future Research}

Examining the exoplanetary detection methodology has shed light on how scientists currently explore the possibility of life elsewhere in our universe. The most prolific methods to date, the radial velocity and transit methods, are applicable to many stellar systems and thus account for the discovery of the majority of exoplanets we know about. Other methods can be used to either discover new exoplanets or corroborate already known exoplanets, but are relatively limited in their applicability. 

Scientists will continue discovering exoplanets in the near future, especially if they use our current techniques in future space missions similar to \textit{Kepler} to look at exoplanets that are relatively close. In addition, collecting more data will improve our search for less detectable exoplanets, especially in the radial velocity method, where larger numbers of observations allows for the detection of smaller velocity amplitudes [Fischer et al, 2015]. Better technology will also help us observe farther planets with higher accuracy; for instance, new adaptive optics technology that corrects for optical wave-front errors caused by Earth’s atmosphere has rectified the pixelation that was once problematic with older spectrometers [Davies and Kasper, 2012]. However, as more of the exoplanets close to us are discovered, it will become increasingly harder to find new exoplanets, as planets that are far away are significantly harder to observe since so few photons are reflected back to the Earth. Hopefully, exoplanetary detection techniques will one day allow us to find out whether live exists elsewhere in our universe.

\section{\label{sec:level1}References}

\end{document}